# Semantic Information G Theory for Range Control with Tradeoff between Purposiveness and Efficiency

Chenguang Lu

*Abstract*—Recent advances in deep learning suggest that we need to maximize and minimize two different kinds of information simultaneously. The Information Max-Min (IMM) method has been used in deep learning, reinforcement learning, and maximum entropy control. Shannon's information rate-distortion function is the theoretical basis of Minimizing Mutual Information (MMI) and data compression, but it is not enough to solve the IMM problem. The author has proposed the semantic information G theory (i.e., Shannon-Lu theory), including the semantic information G measure and the information rate fidelity function $R(G)$ ($R$ is the MMI for the given $G$ of semantic mutual information). The parameter solution of the $R(G)$ function provides a general method to improve the information efficiency, $G/R$. This paper briefly introduces the semantic information G measure and the parametric solution of the $R(G)$ function. Two examples reveal that the parametric solution can help us optimize range control with tradeoff between the purposiveness (i.e., the semantic mutual information) and the information efficiency. It seems that the $R(G)$ function can serve as the theoretical basis of IMM methods, but we still need further research in combination with deep learning, reinforcement learning, and constraint control.

*Keywords*—minimum mutual information, maximum entropy control, semantic information measure, information rate distortion, Shannon Theory generalization.

## I. INTRODUCTION

The recent advances in deep learning, especially the success of AutoEncoders [1], [2], remind us that we need to minimize Shannon's mutual information [3], [4], [5], [6] while maximizing likelihood or estimated mutual information [7], [8]. Therefore, some researchers emphasize that intelligence lies in data compression [9], [10]. Information Theoretic Regularization (ITR) [11] is a typical Information Max-Min (IMM) method. It has been adopted in deep learning [11] and reinforcement learning [12], obtaining good results. ITR uses the difference between two kinds of mutual information as the objective function and minimizes it with the variational method.

IMM is related to the rate distortion theory Shannon initiated [13], [14] and the maximum entropy theory Jaynes proposed [15]. Besides maximum entropy control, people have also proposed minimum entropy [16], [17] and Max-Min entropy control [18]. These are all related to IMM.

However, the rate-distortion theory is insufficient because we also need to maximize some kind of mutual information or minimize some sort of entropy. Therefore, we need a new or updated theory for IMM methods.

I proposed the semantic information G measure 30 years ago [19], [20] (G means the generalization of Shannon's information theory) and then extended the information rate distortion function $R(D)$ to the information rate fidelity function $R(G)$ [21], [22] ($G$ is the lower limit of semantic mutual information). This theory is called the semantic information G theory or Shannon-Lu theory. In the past decade, I have applied the G theory to multilabel classification, maximum mutual information classification, mixed models, and Bayesian confirmation [4], [5], [23]. I have also used the G measure to explain and generalize the maximum entropy method [24]. An important conclusion is that the semantic channel matching with the Shannon channel can maximize semantic mutual information, and the Shannon channel matching with the semantic channel can minimize Shannon's mutual information or maximize information efficiency $G/R$. The G theory might become the fundamental theory supporting IMM methods.

Ranges considered in this article are fuzzy and can be represented by (fuzzy) truth functions or membership functions. Range control also includes error control because the truth function may be equal to $\exp[-d(x, y)]$ [5], [24], where $d(x, y)$ is a distortion function..

This paper aims at 1) demonstrating that the G measure can also be used to evaluate the purposiveness of control results and the semantic information of imperatives; 2) pointing out that semantic communication has an information efficiency issue, just like that energy conversion has a working efficiency issue; 3) showing that using the $R(G)$ function can improve communication efficiencies; 4) simplifying IMM methods.

The main contributions of this article are: 1) demonstrating how to use the $R(G)$ function to optimize maximum entropy control with fuzzy ranges as goals. 2) using the $R(G)$ function's parameter solutions to resolve the problem with the tradeoff between maximizing the purposiveness and maximizing the information efficiency.

## II. THE SEMANTIC INFORMATION G THEORY: AN OUTLINE

### A. The P-T Probability Framework

The semantic information G theory is based on my proposed P-T probability framework [4], [23]. This framework includes two types of probabilities: the statistical denoted by $P$ and the logical by $T$. We define:

1) The $y_j$ is a label or a hypothesis; $y_j(x_i)$ is a proposition. The $\theta_j$ is a fuzzy subset of universe $U$; elements in $\theta_j$ make $y_j$ true. We have $y_j(x) = "x \in \theta_j"$. The $\theta_j$ also means a model or a group of model parameters.

2) A probability defined with "=", such as $P(y_j) = P(Y=y_j)$, is statistical; a probability defined with "$\in$", such as $P(X \in \theta_j)$, is logical. To distinguish them, we define $T(y_j)=T(\theta_j)=P(X \in \theta_j)$ as the logical probability of $y_j$.

3) $T(y_j|x)=T(\theta_j|x)=P(X \in \theta_j|X=x) \in [0, 1]$ is the truth function of $y_j$ and the membership function $m_{\theta_j}(x)$ of $\theta_j$.

The logical probability of a hypothesis $y_j$ may be different from its statistical probability. For example, a tautology's logical probability is 1, whereas its statistical probability is almost 0. We have $P(y_1) + P(y_2) + \ldots + P(y_n)=1$, but there may be $T(y_1) + T(y_2) + \ldots + T(y_n) > 1$.

A semantic channel consists of a group of truth functions: $T(y|x)$: $T(\theta_j|x)$, $j = 1, 2, \ldots, m$, as well as a Shannon channel consists of a group of transition probability functions: $P(y_j|x)$, $j=1, 2, \ldots, n$. According to the above definition, we have:

$$T(y_j) = T(\theta_j) = P(X \in \theta_j) = \sum_i P(x_i)T(\theta_j | x_i). \quad (1)$$

We can put $T(\theta_j|x)$ and $P(x)$ into a Bayes' formula to obtain a likelihood function:

$$P(x|\theta_j) = \frac{T(\theta_j | x)P(x)}{T(\theta_j)}, \quad T(\theta_j) = \sum_i T(\theta_j | x_i)P(x_i). \quad (2)$$

I call (2) the semantic Bayes formula. Since the maximum of $T(y|x)$ is 1, from $P(x)$ and $P(x|\theta_j)$, we obtain:

$$T(\theta_j | x) = \frac{T(\theta_j)P(x | \theta_j)}{P(x)}, \quad T(\theta_j) = 1 \Big/ \max(\frac{P(x | \theta)}{P(x)}). \quad (3)$$

### B. The Semantic Information G Measure

The (amount of) semantic information conveyed by $y_j$ about $x_i$ is defined with log-normalized-likelihood:

$$I(x_i; \theta_j) = \log\frac{P(x_i | \theta_j)}{P(x_i)} = \log\frac{T(\theta_j | x_i)}{T(\theta_j)}. \quad (4)$$

The above formula is illustrated in Fig. 1. We call $I(x_i; \theta_j)$ or its average the semantic information G measure. Fig. 1 indicates that the less the logical probability is, the more information there is; the larger the deviation is, the less information there is; a wrong hypothesis conveys negative information. These conclusions conform to Popper's thoughts [25, p. 294]. $I(x_i; \theta_j)$ is also explained as fidelity (or verisimilitude [25, p. 309]) between $y_j$ and $x_i$.

We call $I(x_i; \theta_j)$ or its average the semantic information G measure. Fig. 1 indicates that the less the logical probability is, the more information there is; the larger the deviation is, the less information there is; a wrong hypothesis conveys negative information. These conclusions conform to Popper's thoughts [25, p. 294]. $I(x_i; \theta_j)$ is also explained as fidelity (or verisimilitude [25, p. 309]) between $y_j$ and $x_i$.

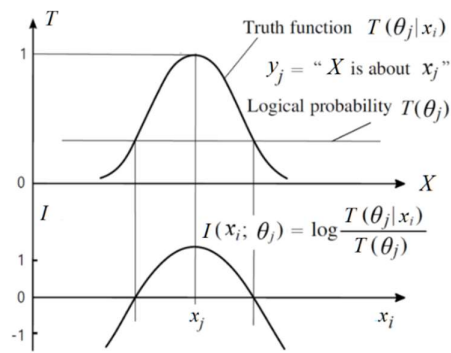

Fig. 1. Semantic information $y_j$ conveys about $x_i$ decreases with the deviation or distortion increasing.

Average semantic information for different $x$ is:

$$I(X;\theta_j) = \sum_i P(x_i | y_j)\log \frac{P(x_i|\theta_j)}{P(x_i)} = \sum_i P(x_i | y_j)\log \frac{T(\theta_j | x_i)}{T(\theta_j)}, \qquad (5)$$

where $P(x|y_j)$ is the sampling distribution. It is easy to prove $I(X; \theta_j) \leq I(X; y_j)$ and that the maximum semantic information criterion is equivalent to the minimum cross entropy and the maximum likelihood criterions [4].

When the sample is enormous so that $P(x|y_j)$ is smooth, we may let $P(x|\theta_j) = P(x|y_j)$ or $T(\theta_j|x) \propto P(y_j|x)$ to obtain the optimized truth function:

$$T^*(\theta_j | x) = \frac{P^*(x|\theta_j)}{P(x)} \Big/ \max_x \left(\frac{P^*(x|\theta)}{P(x)}\right) = \frac{P(x|y_j)}{P(x)} \Big/ \max_x \left(\frac{P(x|y)}{P(x)}\right) = \frac{P(y_j|x)}{\max_x(P(y_j|x))}. \qquad (6)$$

If $P(x|y_j)$ is unsmooth, we may derive a smooth $T^*(\theta_j|x)$ with parameters by using:

$$T^*(\theta_j | x) = \arg\max_{\theta_j} \sum_i P(x_i | y_j)\log \frac{T(\theta_j | x_i)}{T(\theta_j)}. \qquad (7)$$

By averaging $I(X; \theta_j)$ for different $y$, we obtain semantic mutual information:

$$I(X;Y_\theta) = \sum_j \sum_i P(y_j)P(x_i | y_j)\log \frac{P(x_i|\theta_j)}{P(x_i)} = \sum_i \sum_j P(x_i)P(y_j | x_i)\log \frac{T(\theta_j | x_i)}{T(\theta_j)}. \qquad (8)$$

When we fix the Shannon channel and let the semantic channel match the Shannon channel, i.e., let $P(x|\theta_j) = P(x|y_j)$ or $T(\theta_j|x) \propto P(y_j|x)$ for every $j$, $I(X; Y_\theta)$ reaches its maximum, Shannon's mutual information $I(X; Y)$. If we fix the semantic channel $T(\theta_j|x)$ ($j=1,2, …$) and let the Shannon channel match the semantic channel, i.e., let $P(x|y_j) = P(x|\theta_j)$ or $P(y_j|x) \propto T(\theta_j|x)$ as possible for every $j$, we can minimize $I(X; Y)$ or maximize information efficiency $I(X; Y_\theta)/I(X; Y)$.

We define the relationship between the truth function and the distortion function with:

$$d(x, y) = \log[1/T(y|x)] \text{ or } T(y|x) = \exp[-d(x, y)], \qquad (9)$$

where we treat log and exp as a pair of function and anti-function. Letting $d(x, y_j) = (x - x_j)^2 / (2\sigma_j^2)$, we derive:

$$I(X;Y_\theta) = -\sum_j P(y_j)\log T(\theta_j) - E_{P(x_i,y_j)} \frac{(x_i - x_j)^2}{2\sigma_j^2} \qquad (10)$$
$$= I_{max} - \bar{d}.$$

where $\bar{d}$ is the average distortion, and $I_{max}$ is $H(Y_\theta)$.

$I(X; Y_\theta)$ is like regularized squares. Therefore, we can treat the maximum semantic information criterion as a special Regularized Least Squares (RLS) criterion.

## C. The Rate Fidelity Function R(G)

If we change the distortion limit $\bar{d} \leq D$ for the $R(D)$ into $I(X; Y_\theta) \geq G$, then $R(D)$ becomes the information rate fidelity function $R(G)$ [4], [21]. In this case, we replace $d_{ij} = d(x_i, y_j)$ with $I_{ij} = I(x_i; \theta_j)$. Following the derivation of $R(D)$ [14, p. 31], we obtain the parameterized solution [24]:

$$G(s) = \sum_i \sum_j I_{ij} P(x_i)P(y_j | x_i),$$
$$R(s) = sG(s) - \sum_i P(x_i)\log \lambda_i, \qquad (11)$$

$$P(y_j | x_i) = P(y_j)m_{ij}^s / \lambda_i, \; i = 1,2,...; j = 1,2,...$$
$$m_{ij} = T(\theta_j | x_i)/T(\theta_j), \; \lambda_i = \sum_j P(y_j)m_{ij}^s. \qquad (12)$$

To obtain reasonable $P(y)$ and $P(y|x)$, we need iterations by using (12) and (13):

$$P(y_j) = \sum_i P(x_i)P(y_j | x_i). \qquad (13)$$

as we do for the *R(D)* function [26, p. 326].

The shape of any *R(G)* function is a bowl-like curve with the second derivative > 0, as shown in Fig. 2.

For the *R(G)* function, $s = dR/dG$. When $s = 1$, $R$ is equal to $G$. $G/R$ indicates the optimized information efficiency. The *R(G)* function has been applied to image compression according to visual discrimination [22], semantic compression [24], and the EM algorithm convergence proof [4], [27].

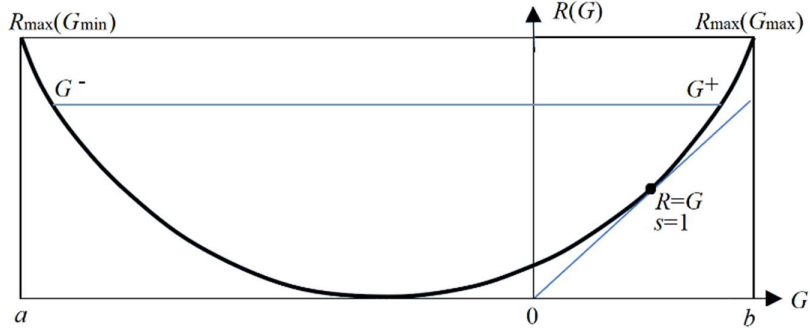

Fig. 2. The information rate fidelity function *R(G)* for binary communication. Any *R(G)* function has a point where *s*=1 and *R* = *G*. For given *R*, there are two anti-functions $G^-(R)$ and $G^+(R)$.

It is worth noting that for given the semantic channel $T(y|x)$, letting $P(y_j|x) \propto T(y_j|x)$ or $P(x|y_j) = P(x|\theta_j)$ does not maximize *G*, but the information efficiency *G/R*. We can maximize both information by (12) with $s \to \infty$.

We can also replace average distortion $\bar{d}$ with fuzzy entropy $H(Y_\theta|X)$ to obtain the rate-truth function $R(\Theta)$ [24]. *R(G)* is more suitable than *R(D)* and $R(\Theta)$ when information is more important than truth.

For given $Y = y_j$, the *R(G)* function is also meaningful. In this case, *R* is the lower limit of $I(X; y_j)$ for given $G = I(X; \theta_j)$; there is no iteration for $P(y)$.

## III. Purposive Information and Optimization

### A. Purposive Information

In my previous studies, the G measure is used to measure the information of labels, predictions, and descriptions. We may call it predictive information because a description can be treated as a special case of a prediction that accords with the fact. Measuring predictive information requires factual tests. For example, we need to use $P(x|y_j)$ to test $P(x|\theta_j)$ for $I(X; \theta_j)$.

We can treat an imperative as a control instruction and need information about whether the control result accords with the control goal. We call this information purposive information. The more accordant the result is, the more information there is. The G measure can also be used to measure the purposive information.

A truth function can represent a control goal. There are some control goals: "The grain yields are close to or more than 5,000 kg/ha", "The death ages of people had better surpass 80 years old", "The cruising distances of electric vehicles had better reach 800 kilometers", and "The error of trains' arrival time should be less than one minute."

We can use the semantic average information formula to measure purposive information:

$$I(X; a_j / \theta_j) = \sum_i P(x_i | a_j) \log \frac{T(\theta_j | x_i)}{T(\theta_j)}, \tag{14}$$

where $\theta_j$ is a fuzzy set, which means that the control goal is a fuzzy range. The $y_j$ in (5) becomes $a_j$ here, indicating the action taken for the control goal $y_j$.

For multi-goal control, the mutual information is

$$I(X; A / \theta) = \sum_j P(a_j) \sum_i P(x_i | a_j) \log \frac{T(\theta_j | x_i)}{T(\theta_j)}, \tag{15}$$

where *A* is a random variable taking a value *a* or $a_j$.

## B. Optimizing Purposive Information

Althoug semantic information formula for imperatives is the same as that for predictive (or descriptive) sentences, the optimization methods are different (see Fig. 3). For predictive sentences, where the fact is fixed, we hope that the predicted range $T(\theta_j|x)$ is consistent with the fact, i.e., $P(x|\theta_j)=P(x|y_j)$, maximizing $I(X; \theta_j)$. For imperatives or range control, we hope that the fact or control result accords with the goal, that is, $P(x|\theta_j)=P(x|y_j)$ or $P(a_j|x) \propto T(\theta_j|x)$, minimizing $I(X; a_j)$ or information efficiency $G/R$.

For multi-goals, we minimize the objective function:

$$f=I(X; A) - sI(X; A/\theta). \tag{16}$$

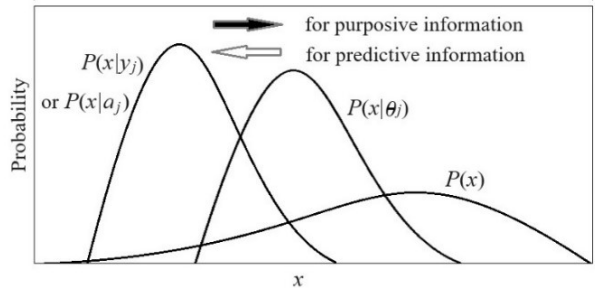

Fig. 3 The ways of optimizing two kinds of semantic information are different. For predictive information, we hope $P(x|\theta_j)$ is close to $P(x|y_j)$ (see the white arrow), whereas for purposive information, we hope $P(x|a_j)$ is close to $P(x|\theta_j)$ (see the black arrow).

When the actual distribution $P(x|a_j)$ approaches $P(x|\theta_j)$, $G/R$ reaches its maximum, 1. To further increase $I(X; a_j/\theta_j)$, we use (12) and (13), replacing $y_j$ with $a_j$. It is worth noting that the many distributions of $P(x|a_j)$ maximize $I(X; a_j/\theta_j)$, but $I(X; a_j)$ is the least as $P(x|a_j) = P(x|\theta_j, s)$.

## IV. RESULTS

### A. Single-goal Optimization for Information and Efficiency

I used the control of adult death ages as an example (see (Fig. 4). Assume the probability distribution $P(x)$ of adults' mortality is normal with $\mu=70$ and $\sigma=10$, without the control. The truth function as the goal is $T(\theta_j|x)=1/[1+\exp[-0.8(x–80)]$. The task is to select an action (a medical policy) leading to a result $P(x|a_j)$ that maximizes $I(X; a_j/\theta_j)$ with less $I(X; a_j)$.

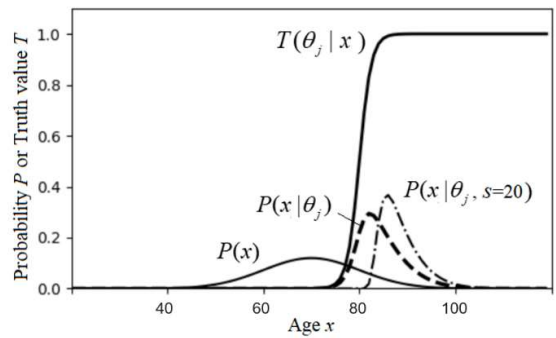

Fig. 4. Adult mortality ages' control. $T(\theta_j|x)$ represents the goal; $P(x|\theta_j) = P(x|\theta_j, s=1)$ is for $G = R$, and $P(x|\theta_j, s = 20)$ makes $G$ close to its maximum.

For this example, $R$ denotes the lower limit of $I(X; a_j)$ for given $G=I(X; a_j/\theta_j)$. We obtained $R(G)$ with $s$ by using (12), as shown in Fig. 5.

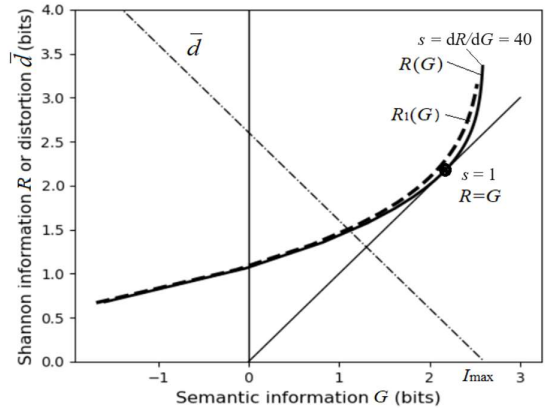

Fig. 5. Two $R(G)$ functions: $R(G)$ was produced by using $P(x|a_j)=P(x|\theta_j, s)$; $R_1(G)$ was produced by using $P(x|a_j) = P(x|\beta_j, s)$, which is a normal distribution; $\bar{d}$ is the average distortion.

The results in Fig. 5 and Table 1 show that information efficiency $R/G$ reaches its maximum, 1, as $P(x|a_j) = P(x|\theta_j, s = 1) = P(x|\theta_j)$. Shannon information $R$ increases faster than semantic information $G^+(R)$ as $s > 1$. $G^+(R)$ almost stops at 2.58 bits as $s > 20$.

The real control result may be a likelihood function $P(x|\beta_j, s)$ different from $P(x|\theta_j, s)$. This paper assumes that we know $P(x|\beta_j, s)$ ($j$=0, 1, …) already without needing to learn them. The main task is to select a proper parameter $\beta_j$.

We obtained function $R_1(G)$ by using $P(x|a_j) = P(x|\beta_j, s)$, a normal distribution. We used the expectation and the square error of $P(x|\theta_j, s)$ as those of $P(x|\beta_j, s)$. Table 1 shows the efficiencies $R/G$ and $R_1/G$ changing with $s$.

TABLE 1. $R(G)$, $R_1(G)$, AND THE EFFICIENCIES CHANGE WITH $s$

| Information (bits) | $s = 1$ | $s = 20$ | $s = 40$ |
|---|---|---|---|
| $R$ | 2.19 | 3.36 | 3.58 |
| $G = G^+(R)$ | 2.19 | 2.58 | 2.59 |
| $G/R$ | 1 | 0.77 | 0.72 |
| $R_1$ | 2.08 | 3.13 | 3.38 |
| $G_1 = G^+(R_1)$ | 1.99 | 2.52 | 2.55 |
| $G_1/R_1$ | 0.95 | 0.80 | 0.76 |

There is always $R < R_1$ for any given $G$, which means $R$ has a higher efficiency than $R_1$. $R_1(G)$ cannot pass through the point at which $R=G$ or $G/R=1$ because $P(x|\beta_j)$ is a symmetrical normal distribution, whereas $P(x|\theta_j)$ is asymmetrical (see Fig. 4). However, $G_1/R_1 = 76$ for $s = 40$ is similar to $G/R=0.77$ for $s = 20$. The similarity means that the normal function is good enough for this example.

One may think that we can select a certain $x$, such as $P(x=80|a_j) = 1$, which can maximize $I(X; a_j/\theta_j)$. However, this selection will make information efficiency be 0.23.

### B. Multi-goal Optimization for Information and Efficiency

Fig. 6 shows a control task with two goals expressed by truth functions $T(\theta_0|x)$ and $T(\theta_1|x)$. We can imagine that the goals are two pastures with fuzzy boundaries, and we need to drive sheep there. If there is no control, the density distribution of the flock is $P(x)$. We need to find a suitable probability distribution $P(a)=P(y)$.

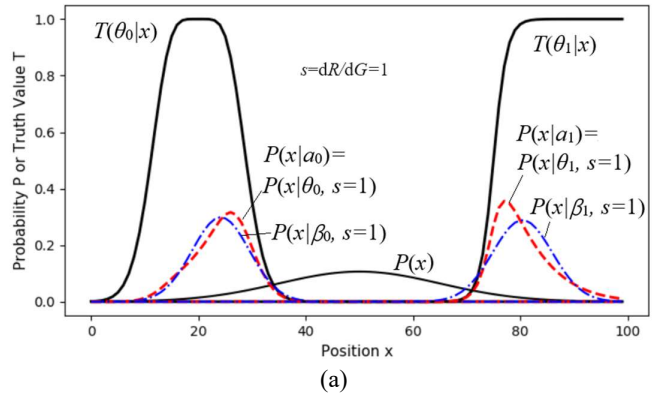

(a)

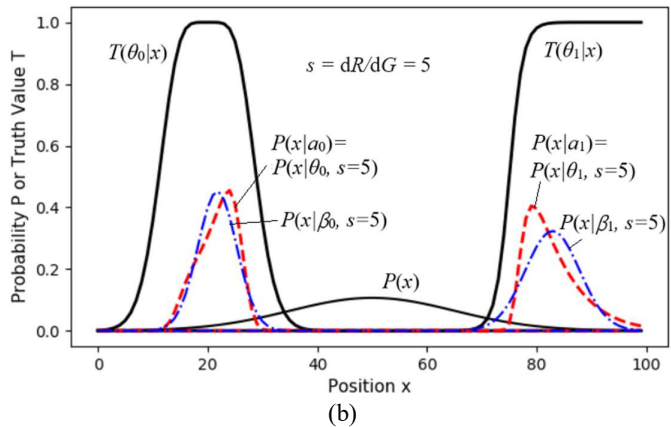

(b)

Fig. 6 $P(x|a_j)=P(x|\theta_j, s)$ (dashed lines) and $P(x|\beta_j, s)$ (dotted lines) in a two-goal control task. (a) is for $s=1$, and (b) is for $s=5$. $P(x|\beta_j, s)$ is a normal distribution produced by action $a_j$ ($j=0,1$).

In Fig. 6, $P(x)$ is a normal distribution ($\mu = 50$, and $\sigma =15$); two truth functions as goals are $T(\theta_0|x) =1-[1-\exp(-(x-20)^2/50)]^3$ and $T(\theta_1|x)=1/[1+\exp(-0.75(x-c))]$. For different $s$, we set initial $P(a_0)=P(a_1)=0.5$. Then we used (12) and (13) with three iterations to obtain $P(a_j|x)$ and $P(a_j)$ ($j=0,1$) and used (12) to get $P(x|a_j)=P(x|\theta_j, s)$. Finally, we adopted (11) to obtain $G(s)$, $R(s)$, and $R(G)$.

Fig.7 displays the $R(G)$ function. Table 2 shows the changes of $P(a)$ with $s$ in (12) and $c$ in $T(\theta_1|x)$.

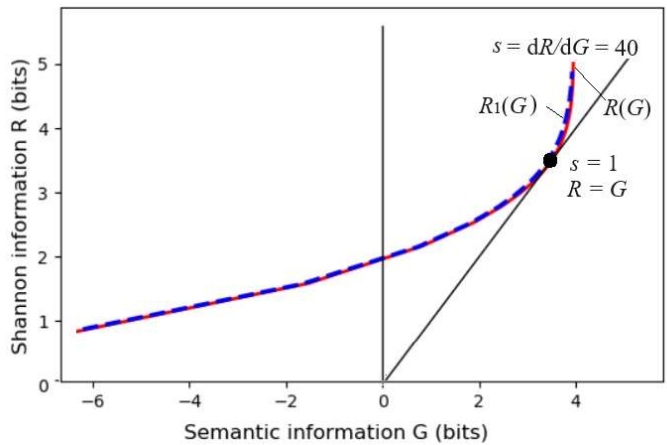

Fig. 7 The $R(G)$ function (solid line) for a two-goal control task. The dashed line represents $R_1(G)$ produced by replacing $P(x|\theta_j, s)$ with a normal function $P(x|\beta_j, s)$ ($j=0,1$).

Table 2 indicates that $G$ and $R$ increase and $G/R$ decreases with $s$ increasing; $P(a_1)$ decreases with $c$ increasing. The change of $P(a)$ reveals that the iterative algorithm can reduce the investment in a difficult goal. Fig. 7 and Table 2 show that when $s$ increases from 5 to 40, $G$ slightly increases, meaning $s=5$ is good enough.

TABLE 2. $R(G)$, $P(a)$ AND $G/R$ CHANGES WITH $S$ AND $C$.

| s | c | $P(a_0)$ | $P(a_1)$ | G (bits) | R(bits) | G/R |
|---|---|---|---|---|---|---|
| 1 | 75 | 0.535 | 0.465 | 3.43 | 3.43 | 1 |
| 1 | 80 | 0.579 | 0.421 | 3.80 | 3.80 | 1 |
| 5 | 75 | 0.540 | 0.460 | 3.89 | 4.29 | 0.907 |
| 5 | 80 | 0.592 | 0.408 | 4.28 | 4.71 | 0.909 |
| 40 | 75 | 0.540 | 0.460 | 3.95 | 5.01 | 0.803 |
| 40 | 80 | 0.592 | 0.408 | 4.33 | 5.34 | 0.811 |

The dashed line for $R_1(G)$ in Fig. 7 indicates that if we replace $P(x|a_j)=P(x|\theta_j, s)$ with a normal distribution, $G$ and $G/R_1$ do not obviously change.

## V. DISCUSSION

### A. How the Results Accord with Theoretic Analyses

Section IV demonstrates that we can optimize constraint control with the help of the $R(G)$ function. Fig. 5, Fig. 7, Table 1, and Table 2 indicate that we can balance between maximizing purposive information and maximizing information efficiency. For example, it is proper to select s between 5 and 20.

Fig. 6 and Table 2 reveal how the iterative algorithm can find a proper distribution $P(a)$ that distributes a smaller proportion on a more difficult goal.

### B. About Alternative Variational Bayes

To obtain $P(y)$ from $P(x)$ and $P(x|\theta_j)$ ($j$=1, 2, …), some people assume $P(y_j)=P(\theta_j)$ ($j$=1, 2, …) and use Variational Bayes (VB) [28]. However, the iterative algorithm for $P(y)$ in (14) and (15) is different and may be called Alternative Variational Bayes (AVB). AVB seems more simpler than VB. Besides the above application, the author has successfully applied AVB to mixture models [4], [27] and semantic compression [24]. Using VB, Friston proposed the Minimum Free Energy Principle [29]. Using AVB, perhaps we can improve this principle to the Maximum Information Efficiency Principle.

### C. Why Do We Use Regularization?

The semantic mutual information formula (10) indicates that error regularization is for more semantic information and a larger likelihood. Section III-B reveals that information regularization (see (16)) is for higher information efficiencies. The above analyses should help us better understand two kinds of regularization.

## VI. CONCLUSIONS

Theoretic analyses and experiments made clear that the semantic information G measure can be used to measure the purposiveness of range control (including error control) and the semantic information of imperatives. This measure can also be used as or for the reward function of reinforcement learning. The parametric solution of the $R(G)$ function provides a general method for the tradeoff between maximizing purposive information $G$ and maximizing the information efficiency $G/R$.

Nevertheless, this paper only focuses on theoretical analyses. We need more experiments in combination with deep and reinforcement learning to verify the theory.